\documentclass[useAMS,usenatbib]{mn2e}
\usepackage{natbib}
\usepackage{graphicx}
\newcommand{\nat}{Nature}
\newcommand{\mnras}{MNRAS}
\newcommand{\apj}{ApJ}
\newcommand{\apjl}{ApJL}
\newcommand{\aap}{A\&A}
\newcommand{\pasj}{Publ.\ Astron.\ Soc.\ Japan}

\newcommand{\eqb}{\begin{equation}}
\newcommand{\eqe}{\end{equation}}
\begin{document}

\title{A model for fast extragalactic radio bursts}
\author[Y.E.Lyubarsky]{Y.E.Lyubarsky\\
Physics Department, Ben-Gurion University, P.O.B. 653, Beer-Sheva
84105, Israel; e-mail: lyub@bgu.ac.il}
\date{Received/Accepted}
\maketitle
\begin{abstract}
Bursts of millisecond duration were recently discovered in the 1 GHz band. There is a strong evidence that they come from $\sim 1 $ Gpc distances, which implies extraordinary high brightness temperature. I propose that these bursts could be attributed to synchrotron maser emission from relativistic, magnetized shocks. At the onset of the magnetar flare, a strongly magnetized pulse is formed, which propagates away through the relativistic magnetar wind and eventually reaches the nebula inflated by the  wind within the surrounding medium. I show that the observed radio bursts could be generated at shocks formed via the interaction of the magnetic pulse with the plasma within the nebula. The model predicts strong millisecond bursts in the TeV band, which could be observed even from distant galaxies.
\end{abstract}

\section{Introduction}
Recently, a number of isolated radio bursts has been discovered in the Parkes pulsar surveys \citep{Lorimer07,Keane11,Thornton13}. These bursts are of millisecond duration and exhibit very large dispersion measure, significantly exceeding the Galactic one. This suggests that the sources are at cosmological distances (an alternative idea of Galactic origin is discussed by \citet{Loeb14} and \citet{Luan14}), $D\sim 1$ Gpc, and the isotropic equivalent of the total emitted energy is as large as ${\cal E}\sim 10^{40}$ erg. The rate of the bursts, $\sim 10^4$ sky$^{-1}$d$^{-1}$, which corresponds to $\sim 10^{-3}$ yr$^{-1}$galaxy$^{-1}$, significantly exceeds that of gamma-ray bursts.

It was suggested, on statistical ground, that the radio bursts could be produced by magnetar hyper flares \citep{PopovPostnov07,PostnovPopov13,Thornton13}. Other proposed progenitors include supernova explosion in a binary containing a neutron star \citep{EgorovPostnov09}, collapse of a supermassive neutron star \citep{FalsckeRezzolla13,Zhang14}, binary white dwarf or neutron star merger \citep{Keane12,Kashiyama13,Lipunov13,Totani13}, evaporation of primordial black holes \citep{Keane12}.

A very short time scale and a large energy release suggest that the source is relativistic. Transparency of the source with respect to the induced scattering of the radio emission implies that the bulk Lorentz factor exceeds a few thousand \citep{Lyubarsky08}. Even if the source is highly relativistic, the inferred brightness temperature is very high therefore one has to look for a coherent emission mechanism. In such cases, it is a common practice to appeal to charged "bunches" \citep{FalsckeRezzolla13,Kashiyama13,Katz13}. The problem with these models is that the very existence of "bunches" with the required properties is just postulated. Alternatively, the high brightness temperature emission could be attributed to maser mechanisms, which of course could be formulated (optionally) in terms of "bunches" but in this case, the bunches are naturally formed via interaction with the radiation field itself so that a self-consistent model could be developed.

\citet{SagivWaxman02} considered synchrotron maser radio emission from gamma-ray bursts. They assumed that the required inverse population of relativistic electrons is formed due to spontaneous synchrotron energy losses. However, the validity of this assumption at real conditions has not been checked.
Earlier \citet{UsovKatz00} suggested that if gamma-ray bursts are produced by relativistic, strongly magnetized winds, they may be accompanied by short pulses of low-frequency radio emission generated by strong coherent oscillations of the current separating the magnetized wind from the non-magnetized plasma incident on the wind front. These oscillations arise due  to the synchrotron instability of a proton half-circle formed by the reflecting particles. However, the incident plasma should be separated from the wind by a forward shock formed due to the Weibel instability of the reflected particles; then the current at the contact discontinuity between the magnetized wind and the shocked upstream plasma is stable and no coherent emission is expected.

A powerful synchrotron maser emission is generated at the front of a relativistic shock provided the upstream plasma is magnetized \citep{LangdonAronsMax88,Hoshino92,Gallant92}. The necessary magnetization is not large, the ratio of the magnetic to the plasma energy should be $\sigma>10^{-3}$ \citep{SironiSpitkovsky11}. In this case, the shock transition is mediated by Larmor rotation of the incoming particles so that a ring distribution in the momentum space is formed, which is relaxed via stimulated emission of low-frequency electro-magnetic waves.   In this Letter, I show that the cosmological radio bursts may be attributed to the maser emission from relativistic shocks produced by magnetar hyper flares. 

\section{A model}
According to the standard model of the soft gamma repeaters \citep{ThompsonDuncan95},
the flare occurs in the magnetosphere of the magnetar when the
magnetic tension within the star breaks the crust thus opening
the way to a new equilibrium. The magnetic field within the star is untwisted, the relaxation time, $\sim 0.5$ s, being determined by the relatively low Alfven velocity within the star. An extreme case of magnetar's flare could occur when the neutron star itself eventually becomes unstable to a dynamic overturning instability that destroys most of its dipole moment in a single event \citep{Eichler02}. During a hyper flare, a significant fraction  of the magnetic energy, $B_*^2R_*^3=10^{48}B_{*15}^2$ erg, is released
producing the so called hard spike with the duration $0.3-0.5$ s.
In the course of the relaxation process, slow motions of the neutron star crust distort the structure of the abovelying magnetic field producing unstable configurations, which explode releasing the magnetic energy.
Inasmuch as the Alfven velocity in the magnetosphere is close to the speed of light, the magnetospheric explosions are very fast, which is confirmed by the short observed
rise time of the flare, $\sim 1$ ms \citep{Palmer05}.
This suggests that the hard spike is in fact formed by
many fast bursts merged into a single long flare.

The energy released during the hard spike is too high to be confined by the magnetar magnetic field therefore an outflow is formed. Observations of the
expanded radio nebula produced by the hyper flare from SGR
1806-20 \citep{Gaensler05,Gelfand05,Granot05,Taylor05} show that the outflow is only mildly relativistic therefore it could not give rise to a relativistic shock.  However, one can speculate that at the onset of the burst, a strong magnetic perturbation produces strong MHD waves (Alfven and magnetosonic), which propagate outwards in the non-perturbed magnetosphere. These waves open the magnetosphere sweeping the magnetic field lines into a narrow pulse propagating outwards  through the magnetized magnetar wind.
The amplitude of the pulse may be presented as
\eqb
B_{\rm pulse}=bB_*\frac{R_*}{r}=10^5\frac{bB_{*15}}{r_{16}}\, \rm G,
\label{B_pulse}\eqe
where $r$ is the distance from the magnetar, $R_*$ the radius of the star, $B_*$ the surface magnetic field, $b<1$ the dimensionless constant. Here and thereafter, the notation $A=10^xA_x$ in cgs units is used.
In this paper, I show that a short radio burst may be generated when the pulse reaches the nebula inflated by the magnetar wind in the surrounding gas.

The magnetar wind resembles the pulsar wind. It is formed by magnetized electron-positron plasma ejected from the light cylinder. The wind power is determined by the spin-down luminosity:
\eqb
L_{\rm sd}=\frac{B_*^2R_*^6\Omega^4}{c^3}=4\cdot 10^{34}B_{*15}^2\Omega^4\,\rm erg\, s^{-1},
\label{Lsd}\eqe
where  $\Omega$ is angular velocity of the magnetar.  Since typical magnetar periods are in the range 5-8 s, one can conveniently normalize $\Omega$ by unity. Inasmuch as the characteristic life time of magnetars is only ~1000 years, they remain within the supernova ejecta.
The magnetar wind is terminated when the wind bulk pressure, $L_{\rm sd}/4\pi r^2c$, is balanced by the confining pressure, $p$. Beyond the termination shock, the shocked plasma of the wind inflates a bubble resembling a pulsar wind nebula (PWN).  The radius of the wind termination shock is estimated as
\eqb
r_s=\sqrt{\frac{B_*^2R_*^6\Omega^4}{4\pi pc^4}}=3.1\cdot 10^{15}\frac{B_{*15}\Omega^2}{p^{1/2}_{-8}}\,\rm cm.
\label{shock_radius}\eqe
I take the fiducial pressure within the nebula $p=10^{-8}$ dyn cm$^{-2}$, which corresponds to a nebula expanding with the velocity 500 km s$^{-1}$ within a medium with the density $4\cdot 10^{-24}$ g cm$^{-3}$.
One can expect that, like PWNe, the magnetar nebula is filled with relativistic electron-positron plasma and magnetic fields roughly at the equipartition.

When the electro-magnetic pulse  arrives at the wind termination shock, it pushes the plasma outwards like a magnetic piston. A forward shock propagates through the magnetized, relativistically hot plasma of the nebula whereas a reverse shock goes into the magnetic piston. Between the shocks, a contact discontinuity separates the shocked plasma of the nebula from the magnetic piston.  At the contact discontinuity, the magnetic pressure of the pulse is balanced by the bulk pressure of the relativistically hot plasma entering the forward shock.

The reverse shock is weak because the piston is highly magnetized. Therefore  the magnetic field in the piston is nearly the same on both sides of the shock and may be  estimated by substituting the distance (\ref{shock_radius}) into equation (\ref{B_pulse}):
\eqb
B_{\rm pulse}=3.2\cdot 10^5\frac{bp^{1/2}_{-8}}{\Omega^2}\,\rm G.
\label{B}\eqe
Let the contact discontinuity, as well as the plasma downward of both the forward and the reverse shocks, move with respect to the lab frame with the Lorentz factor $\Gamma_{\rm cd}$. The pressure balance in the contact discontinuity frame is written as
\eqb
\frac{B_{\rm pulse}^2}{8\pi\Gamma_{\rm cd}^2}=4\xi p\Gamma_{\rm cd}^2,
\label{pressure_balance}\eqe
where the dimensionless factor $\xi<1$ takes into account that  most of the energy in the nebula is contained in high energy particles, which could loose their energy before completing the full turn in the downstream magnetic field. In this case, the bulk pressure decreases. This issue will be elaborated in section 4.

Substituting equation (\ref{B}) into equation (\ref{pressure_balance}), one finds
\eqb
\Gamma_{\rm cd}=
1.8\cdot 10^4\frac{b^{1/2}}{\xi^{1/4}\Omega}.
\label{Gamma_cd}\eqe
The piston pushes forward the plasma and heats it.  When the piston passes the distance $\Delta r$ within the nebula, the $pdV$ work on the plasma is $B^2_{\rm pulse}\Delta r/(8\pi\Gamma_{\rm cd}^2)$ per unit square. The energy in the pulse  per unit square is  $\sim (B^2_{\rm pulse}/4\pi)l$, where $l$ is the pulse width. This energy is totally transferred to the plasma at the distance
\eqb
\Delta r=2\Gamma^2_{\rm cd}l=6.4\cdot 10^{14}\frac{bl_6}{\xi^{1/2}\Omega^2}\,\rm cm.
\label{Delta_r}\eqe
One sees that the pulse does not penetrate deep into the nebula.

The reverse shock is strongly magnetized; the forward shock is moderately magnetized (recall that in PWNe, the magnetic field is roughly in the equipartition with the plasma). Therefore both the reverse and the forward shocks are mediated by Larmor rotation. Let us consider emission from these shocks. 

\section{Maser emission from the forward shock}
Taking into account that the magnetization of the plasma within the nebula is not small and only increases at the shock front, let us assume for simplicity that in the shocked plasma the magnetic field is in equipartition, which means that the magnetic field is $B=2^{-1/2}B_{\rm pulse}$.
In the downstream plasma, the inverse population is formed at the particle energies less than $m_ec^2\Gamma_{\rm cd}$. In this case, the maser emission predominantly occurs at the rotation frequency of these particles\footnote{More exactly, the emission peaks at the frequency $\Omega'_B$ if $\sigma>1$. In the opposite case, the radiation peak occurs at a larger frequency, $\omega=0.5\sigma^{-1/4}\Omega'_B$ \citep{Lyubarsky06}. Taking into account that in our case, $\sigma\sim 1$, I neglect this difference in rough estimates. }, $\nu'=\Omega'_B/2\pi=eB'/(2\pi m_ec\Gamma_{\rm cd})$, where the quantities in the downstream frame are marked by prime. In the observer's frame, the main radiation frequency is estimated as
\eqb
\nu_{0}=\frac 1{\sqrt{2}}\frac{eB_{\rm pulse}}{2\pi m_ec\Gamma_{\rm cd}}=42\frac{\xi^{1/4}b^{1/2}p_{-8}^{1/2}}{\Omega}\,\rm MHz.
\label{nu0}\eqe
This frequency is lower than the observed one but simulations show that the emitted spectrum extends to higher frequencies \citep{Gallant92}. This is presumably because the particles radiate away their energy and become to rotate faster.

If a fraction $\eta$ of the particle energy is radiated away, the observed luminosity is estimated as
\eqb
L=\eta \dot{N}m_ec^2\Gamma_{\rm cd}^2,
\eqe
where $\dot{N}$ is the number of particles entering the shock per unit time,
\eqb
\dot{N}=4\pi r^2_s c n,
\eqe
$n$ the particle density in the nebula.

The radiation power reaches a few per cent of the upstream energy \citep{Gallant92} however, most of the energy is emitted at the basic frequency (\ref{nu0}). The fraction of the energy emitted at the observation frequency, $\sim 1$ GHz, is smaller therefore we normalize $\eta$ by $10^{-3}$. The pulse penetrates into the nebula only the distance $\Delta r$ given by equation (\ref{Delta_r}). 
Therefore the total isotropic energy of the maser radio emission is estimated as
\eqb
{\cal E}_{\rm radio}=\eta 4\pi r_s^2m_ec^2n\Gamma_{\rm cd}^2\Delta r
=2\cdot 10^{46}\eta_{-3}b^2B^2_{*15}l_6\xi^{-1}n\,\rm erg.
\eqe
The energies of the observed radio bursts, $\sim 10^{40}$ erg, could be achieved at the particle densities in the nebula  $n\sim 10^{-6}$ cm$^{-3}$. Note that this is comparable with the particle density in the Crab nebula.
Taking into account the transit time effects, the observed duration of the burst is
\eqb
\delta t=\frac 2{c\Gamma_{\rm cd}^2}\Delta r
=1.3\cdot 10^{-4}l_6\,\rm s.
\label{duration}\eqe

\section{Very high energy emission from the forward shock}
In PWNe, the energy density and the pressure is determined by high energy particles with the Lorentz factors $\gamma_{E}\sim 10^4-10^6$. At $\gamma<\gamma_E$  the particle spectrum is a power law, $\propto\gamma^{-\alpha}$, with a rather shallow slope, $\alpha=1-1.5$. With such a slope, most of particles find themselves at low energies whereas the plasma pressure and the energy density are determined by a small fraction of high energy particles. Normalizing by the plasma pressure, $p$, one can write the particle spectrum  as
\eqb
\frac{dn_{\rm HE}}{d\gamma}=\frac{(2-\alpha)p}{3m_ec^2\gamma_E^2}\left(\frac{\gamma_E}{\gamma}\right)^{\alpha}.
\label{particle_spectrum}\eqe

Downstream of the forward shock, these particles rotate in the magnetic field. In the downstream frame, the particle  Lorentz factor is $\gamma'=2\gamma\Gamma_{\rm cd}$ and the particle momentum is transferred to the downstream medium if it completes at least a quarter of the Larmor rotation before it looses the energy to the synchrotron emission. The corresponding condition is
\eqb
\frac 49\omega'_B\gamma'^2<m_ec^3/e^2,
\label{cond_syncr}\eqe
where $\omega'_B=eB_{\rm pulse}/({\sqrt{2}m_ec\Gamma_{\rm cd}})$
is the Larmor frequency in the downstream frame.
This condition means that the maximal energy of synchrotron photons detected by an observer at rest is
\eqb
\varepsilon_1\sim\frac{\hbar m_ec^3}{e^2}\Gamma_{\rm cd}=1.3\frac{b^{1/2}}{\xi^{1/4}\Omega} \,\rm TeV.
\label{UHEband}\eqe

Substituting the estimates (\ref{B}) and (\ref{Gamma_cd}) into equation (\ref{cond_syncr}), one finds that only particles with the Lorentz factors
\eqb
\gamma<\gamma_1=9.2\cdot 10^4\frac{\xi^{1/8}\Omega^{3/2}}{b^{3/4}p_{-8}^{1/4}}
\label{gamma1}\eqe
transfer their momenta to the downstream medium.

The coefficient $\xi$ was introduced in equation (\ref{pressure_balance}) in order to take into account that because the most energetic particles loose their energy before completing a Larmor turn, only a fraction $\xi<1$ of the total bulk pressure is really exerted on the piston. The coefficient $\xi$ is in fact the ratio of the energy contained in the particles with the Lorentz factors $\gamma<\gamma_1$ to the total plasma energy. For the particle distribution function (\ref{particle_spectrum}), one can write
\eqb
\xi=\frac 3p\int_1^{\gamma_1}m_ec^2\gamma \frac{dn_{\rm HE}}{d\gamma}d\gamma=\left\{\begin{array}{ll} 1 & \gamma_1>\gamma_E \\ \left(\frac{\gamma_1}{\gamma_E}\right)^{2-\alpha} & \gamma_1<\gamma_E
\end{array}\right.
\eqe
Substituting equation (\ref{gamma1}), one gets an equation for $\xi$. At $\alpha=1$, this yields
\eqb
\xi=\min \left(1;\,\frac{1.1\Omega^{12/7}}{\gamma_{E5}^{8/7}b^{6/7}p_{-8}^{2/7}}\right).
\eqe
For $\alpha=1.5$ one gets
\eqb
\xi=\min\left(1;\,\frac{\Omega^{4/5}}{\gamma_{E5}^{8/15}b^{2/5}p_{-8}^{2/15}}\right).
\eqe

In any case the total available energy,
\eqb
{\cal E}_{\rm total}=B_{\rm pulse}^2r^2l=10^{48}b^2B_{*15}^2l_6
\eqe
is emitted in the ultra high energy band (\ref{UHEband}) as a pulse with the duration (\ref{duration}).
Such a pulse could be observed from the distance $\sim 100$ Mpc.

\section{Maser emission from the reverse shock}
Let us now consider the maser emission from the reverse shock. Let the plasma within the magnetic pulse move with the Lorentz factor (estimated below) $\Gamma_{\rm pulse}$. Downstream of the reverse shock, the particles rotate in the magnetic field with the Lorentz factor
\eqb
\gamma'=\Gamma_{\rm pulse}/(2\Gamma_{\rm cd}).
 \eqe
Just as in the case of the forward shock, one can see that the maser emission is generated at the frequencies
\eqb
\nu>\nu_0=\frac{eB_{\rm pulse}}{2\pi m_ec\gamma'}.
\eqe
If the pulse carries $N$ particles, and the radiation efficiency is $\eta$ (see section 3), the total emitted energy is
\eqb
{\cal E}=\eta Nm_ec^2\Gamma_{\rm pulse}.
\eqe

The magnetic pulse picks up the plasma accumulated in the magnetosphere. At the quiescent state, the magnetar magnetosphere is filled by electron-positron plasma generated in cascades induced near the surface of the star by slow untwisting of magnetospheric field lines \citep{ThompsonLyutikovKulkarni02, BeloborodovThompson07,Thompson08,Beloborodov13a}. The pairs going beyond the distance $\sim 10 R_*$
could not come back to the star because the X-ray emission from the surface exerts on them strong pressure in the cyclotron resonance. These pairs are accumulated in the upper magnetosphere until the injection is balanced by annihilation \citep{Beloborodov13b}. Most of pairs is accumulated in the outer parts of the magnetosphere, $r\sim c/\Omega$. Let us estimate the total number of the accumulated pairs.

The pair injection rate is determined by currents in the magnetosphere. Let us consider a twisted bundle of magnetic field lines with apex radii $r_{\rm max}\sim c/\Omega$. A cascade near the surface of the star injects the pairs, from two poles, at the rate
\eqb
\dot{N}=2\kappa I/e,
\eqe
where $I$ is the electric current in the bundle, $\kappa$ the pair production multiplicity. The theory of pair production in magnetars predicts  $\kappa\sim 100$ \citep{Beloborodov13a}. The current in the light cylinder region could not exceed the Goldreich-Julian current flowing in the open field line tube, $I_{GJ}=B_*\Omega^2R_*^3/2c$, because the "toroidal" component of the magnetic field could not exceed, by stability considerations, the poloidal one. Therefore the particle injection rate (from two poles) into the external part of the closed magnetosphere is roughly the same as the rate of particle ejection from the magnetosphere into the wind zone:
 \eqb
 \dot{N}=\kappa\frac{B_*\Omega^2R_*^3}{ec}=7\cdot 10^{33}\kappa_2B_{*15}\Omega^2\,\rm s^{-1}.
 \label{injection}\eqe

The pair density in the magnetosphere, $n$, is established when the injection is balanced by annihilation:
\eqb
\dot{N}=\langle\sigma v\rangle n^2 V.
\eqe
Here  $V\sim (c/\Omega)^3$ is the volume of the magnetosphere, $\langle\sigma v\rangle$ the annihilation rate. The pairs in the upper magnetosphere are non-relativistic due to the Compton interaction  with the X-ray emission from the star's surface; then $\langle\sigma v\rangle=\pi r_e^2 c$.  Now the total number of accumulated pairs is estimated as
\eqb
N=nV=5.1\cdot 10^{39}\sqrt{\frac{\kappa_2B_{*15}}{\Omega}}.
\label{N}\eqe

The Lorentz factor of the plasma  within the pulse may be expressed via that in the wind frame, $\Gamma'_{\rm pulse}$,  and the Lorentz factor of the wind, $\Gamma'_{\rm wind}$, as
\eqb
\Gamma_{\rm pulse}=2\Gamma_{\rm wind}\Gamma'_{\rm pulse}.
 \eqe
Let the pulse be fully electro-magnetic so that the electric and the magnetic fields of the pulse are equal, $E_{\rm pulse}=B_{\rm pulse}$ and the pulse propagates with the speed of light. The plasma within the pulse moves with the velocity $v=cE/B$.

In the wind frame, the plasma moves with the velocity
\eqb
v'_{\rm pulse}=\frac{E_{\rm pulse}}{B_{\rm pulse}+B_{\rm wind}}c,
\label{v'}\eqe
where
\eqb
B_{\rm wind}=B_*\frac{R_*^3\Omega^2}{c^2r}
\eqe
is the magnetic field in the wind. Note that the rhs of equation (\ref{v'}) is independent of both the frame of reference and the distance, $r$. Taking into account that $B_{\rm pulse}\gg B_{\rm wind}$, one finds the Lorentz factor of the plasma in the pulse as
\eqb
\Gamma'_{\rm pulse}=\sqrt{\frac{B_{\rm pulse}}{2B_{\rm wind}}}=\sqrt{\frac b2}\frac{c}{\Omega R_*}=
2.1\cdot 10^4\frac{b^{1/2}}{\Omega}.
\eqe

Now let us consider the Lorentz factor of the wind.
Without dissipation, the wind practically stops accelerating after passing the fast magnetosonic velocity so that the final Lorentz factor is $\Gamma_{\rm wind}= {\rm  few}\times [L_{sd}/(m_2c^2\dot{N})]^{1/3}$ \citep{Beskin98}.
Substituting equations (\ref{Lsd}) and (\ref{injection}), one finds
\eqb
\Gamma_{\rm wind}={\rm few}\times 100 (B_{*15}\Omega^2/\kappa_2)^{1/3}.
\label{Gamma_wind}\eqe
The wind could be accelerated further out due to dissipation of alternating magnetic fields \citep{LyubarskyKirk01,KirkSkjaeraasen03}. The field is dissipated in current sheets separating the stripes of opposite magnetic polarity, the distance between the stripes being $d=\pi c/\Omega=9\cdot 10^{10}/\Omega$ cm. One sees that at the Lorentz factor (\ref{Gamma_wind}), the adjacent stripes remain causally disconnected when the wind arrives at the termination shock, which implies that dissipation in the current sheets could hardly accelerate the flow. Taking this into account, I normalize the wind Lorentz factor by 1000.

Now the particles Lorentz factor in the downstream frame is presented as
\eqb
\gamma'=\frac{\Gamma_{\rm wind}\Gamma'_{\rm pulse}}{\Gamma_{\rm cd}}=1.2\cdot 10^3\Gamma_{\rm wind, 3}.
\label{gamma'}\eqe
The main radiation frequency for the maser emission is
\eqb
\nu_0=750\frac{bp_{-8}^{1/2}}{\Gamma_{\rm wind,3}\Omega^2}\,\rm MHz.
\eqe
Note that this frequency is close to the observed one. The radiation efficiency, $\eta$, at $\nu\sim\nu_0$ may be as high as a few per cent \citep{Gallant92}. The total emitted energy is  estimated as
\eqb
{\cal E}=1.8\cdot 10^{39}\eta_{-2}\Gamma_{\rm wind,3}\sqrt{\frac{b\kappa_2B_{*15}}{\Omega^{3}}}\,\rm erg.
\eqe
One sees that the reverse shock could produce the observed radio bursts`     as well as the forward shock.

\section{Conclusions}
It has already been proposed on statistical ground that cosmological fast radio bursts are produced by magnetar hyper flares \citep{PopovPostnov07,PostnovPopov13,Thornton13}. In this Letter, I develop a model for such bursts. Magnetars, like pulsars, permanently emit relativistic, magnetized winds inflating in the surrounding medium nebulae filled with the relativistic electron-positron pairs and magnetic fields. The magnetized electron-positron plasma is injected into the nebula at the wind termination shock that occurs at the distances $\sim 10^{15}-10^{16}$ cm from the magnetar.  When the magnetosphere of the magnetar is violently restructured giving rise to a gamma flare, a strong magnetic pulse propagates outwards in the magnetar wind and eventually crosses the termination shock. Within the nebula, the pulse pushes the plasma outwards producing a strong, highly relativistic forward shock whereas  within the pulse itself, a reverse shock arises.

Both the forward and the reverse shocks are magnetized, i.e. mediated by the Larmor rotation. At the front of such shocks, the ring-like particle distribution gives rise to powerful syncrotron maser emission  \citep{LangdonAronsMax88,Hoshino92,Gallant92}. Here I have estimated the parameters of this emission and have shown that it could naturally account for the observed fast radio bursts.

One can expect that in magnetar nebulae, like in PWNe, most of the energy is contained in highly relativistic pairs.
At the forward shock, they are boosted to very high Lorentz factors so that in the shock enhanced magnetic field, they immediately loose the acquired energy to synchrotron emission. The characteristic frequency of this emission is Lorentz shifted into the TeV band. Therefore the model predicts that magnetar hyper flares are accompanied by millisecond very-high energy gamma-ray bursts, which could be observed by Cherenkov telescopes from the distances $\sim 100$ Mpc.

\end{document}